\documentclass[12pt]{article}

\usepackage{fullpage}
\usepackage{graphicx}
\usepackage{graphics}
\usepackage{epsfig}
\usepackage{latexsym}
\usepackage{amsfonts}
\usepackage{amssymb}
\usepackage{amsmath}

\title{Condensation transitions in a model for a directed network with
weighted links}

\author{A. G. Angel, T. Hanney, M. R. Evans \\  {\small
      School of Physics, University of Edinburgh, Mayfield Road,
      Edinburgh, EH9 3JZ, UK} } 

\begin{document}

\maketitle

\begin{abstract}
An exactly solvable model for the rewiring dynamics of weighted,
directed networks is introduced. Simulations indicate that the model
exhibits two types of condensation: (i) a phase in which, for each
node, a finite fraction of its total out-strength condenses onto a
single link; (ii) a phase in which a finite fraction of the total
weight in the system is directed into a single node.  A virtue of the
model is that its dynamics can be mapped onto those of a zero-range
process with many species of interacting particles --- an exactly
solvable model of particles hopping between the sites of a
lattice. This mapping, which is described in detail, guides the
analysis of the steady state of the network model and leads to
theoretical predictions for the conditions under which the different
types of condensation may be observed. A further advantage of the
mapping is that, by exploiting what is known about exactly solvable
generalisations of the zero-range process, one can infer a number of
generalisations of the network model and dynamics which remain exactly
solvable.
\end{abstract}

\section{Introduction} \label{introduction}
Networks, and in particular weighted networks, have a long history in
the social, natural and engineering sciences. The many and varied
examples of networks range from systems such as transportation networks to
ecological, biochemical and social networks \cite{AB02,DM03}.
Basically, a network is a graph containing nodes connected by links
(or edges) --- the network is said to be `weighted' when weights are
assigned to the links. The interest in networks within the physics
community was initially with regard to the statistical properties of
the connectivity structure of unweighted networks (i.e., networks
where links are restricted to have weight one or zero only, indicating
a connection or no connection respectively). Studies focussed, for
example, on the degree distribution, i.e., the distribution of the
number of links connected to a node, and the clustering coefficient
which probes how trios of nodes are correlated.  Subsequently simple
models for growing networks exhibiting interesting statistical
properties were proposed \cite{BA99}.

Recently, weighted networks have also become of interest to this
community. These networks may be defined through their adjacency
matrix $w_{ij}$, which gives the weight of the link from node $i$ to
node $j$. The links may be directed or undirected corresponding to an
asymmetric or symmetric adjacency matrix.  Data are available for many
real weighted networks, of either directed or undirected kind, leading
to studies of the statistical properties of Scientific Collaboration
Networks (SCN) \cite{Newman01a,Newman04} and the World-wide Airport
Network (WAN) \cite{GA04,BBPV05}. Typical quantities used to describe
the topology of general weighted networks are: the distribution of
in-strength and out-strength, which are defined as the total weight
going into or out of a node due to all the connected links; weighted
clustering coefficients; generalisations of the idea of minimal
spanning trees to weighted networks \cite{NR02,MAB04}.  An interesting
aspect of weighted networks is how the weights of the links are
correlated to the topology of the network: the weights may be
determined solely by the global topology, they may be independent, or
they may be correlated in some more complicated way.  For example, in
Transportation Networks, where the weight represents the flow through
a link, the weights are determined solely by the topology; in the SCN
it appears that the degree of a node and weights of its links are
uncorrelated; in the WAN, correlations between the node degree and the
weights of its links lead the strength to grow faster than the degree.

Attention has recently focussed on constructing simple models of
`growing' weighted networks, with the aim being to investigate the
resulting degree and weight distributions and the relation between the
two. Initial work had the weight of links determined by the degree of
the node they were attached to \cite{MAB04,YJBT01,AK05,AKR05}.  To
loosen this coupling between weight and degree, a model with dynamical
evolution of weights during growth was introduced in
\cite{BBV04,BBV04a, BBV04b}.  Moreover, models of growing networks in
which the dynamics depend on the weights of the links rather than the
degree of the nodes were proposed in \cite{DM04}.  For a review of
recent work on weighted networks see \cite{BBPV05}.

An alternative class of evolving network models, which is the focus of
this work, is that of `rewiring' or `equilibrium' networks
\cite{DM03,BCK01,DMS03}. In these models, usually considered for
unweighted networks, the number of nodes is fixed and the dynamics
involves the rewiring of links between nodes, although the effect of
creating and destroying links has been examined recently
\cite{AELM05}. For rewiring dynamics, one is interested in the
steady-state properties of an ensemble of networks.  Under certain
conditions on the rewiring rules \cite{EH05} one can obtain a
condensed phase where one node (the condensate) attracts a finite
fraction of the available links. Thus there is a single extensively
connected node while the other nodes exhibit a power-law background in
the degree distribution. Hence the network is scale free if one
discounts the condensate. The transition is reminiscent of the
`gelation' transition in growing networks from the scale free phase to
a phase with a dominant `hub' \cite{KRL00,KR01,BB01}.

These rewiring networks may be related to interacting particle systems
such as the zero-range process \cite{Evans00} for which condensation
transitions have been widely studied --- for a recent review see
\cite{EH05}. In the zero-range process, particles hop between the
sites of a lattice with a rate that depends on the number of particles
at the site of departure. The idea behind the mapping is to identify
the nodes of the network with the sites of the lattice model and the
links of the network become the particles --- the rewiring of links
corresponds to particles hopping from site to site. At the simplest
level, this mapping is approximate in that the number of particles at
a site represents only the degree of the corresponding node, but does
not determine the other nodes to which a node is attached. However,
the steady states of the two systems have the same form.

The aim of the present work is to introduce a model for the rewiring
dynamics of a directed, weighted network. By virtue of an exact
mapping between the network model and a set of coupled interacting
particle systems, we are able to analyse exactly conditions on the
rewiring rates under which the network will exist in some kind of
condensed phase. Specifically, we define the network model in Section
\ref{network model}, the key features of which are:
\begin{enumerate}
\item The network is directed.

\item The out-strength of each node is conserved under the dynamics
but the in-strength evolves.

\item The dynamics are governed by the weights of links and
the in-strength of the nodes.

\end{enumerate}
In addition, the weights are integer variables, although this
constraint may be relaxed (as we discuss in Section
\ref{generalisations}). Thus we define a weighted network model with a
fixed number of nodes where strong nodes and links with large weight
may tend to attract more weight.

In Section \ref{mapping}, we show how this model enjoys a mapping to
an interacting particle system known as a multi-species zero-range
process \cite{EH03,HE04}. The idea is that, since the out-strength of
each node is conserved, the dynamics at each node is basically the
exchange of weight between the links coming out of that node. Thus
each node may be related to an interacting particle system in which
the exchange of particles between sites represents the exchange of
weight between links. However, since the dynamics at each node also
depends on the in-strength, these interacting particle systems are
coupled.  This mapping of the weighted network to a set of coupled
interacting particle systems contrasts with the previous mappings of
an unweighted network to a single interacting particle system
discussed in \cite{DM03,BCK01,DMS03,EH05}.

The results of simulations, presented in Section \ref{simulations},
indicate two distinct types of condensation within this model: the
first is when, for any given node, a finite fraction of the
out-strength condenses onto one weight out of that node; the second is
when a finite fraction of the total weight in the system is directed
onto one particular node.  The first scenario gives a realisation of a
transition in the `disparity', the idea of `disparity', discussed in
the context of internet traffic \cite{BGG03} and simple growing
weighted network models \cite{B05}, being that one weight from a node
is dominant over the others from that node. The second scenario is
related to the usual condensation transition occurring in unweighted
rewiring networks, wherein a single `hub' acquires a finite fraction
of the total number of links in the system.

We exploit the mapping to the interacting particle system in Section
\ref{theory}, in order to solve and analyse steady state properties of
the network model exactly. In particular, we derive conditions on the
rewiring rates which lead to the two types of condensation observed in
simulations. The mapping also enables one to identify a number of
possible generalisations of the network model and its rewiring
dynamics, which we discuss in Section \ref{generalisations}. Finally,
we conclude in Section \ref{conclusion}.

\section{Model}
\label{network model}

\begin{figure}
\begin{center}
\begin{equation}
\setlength{\arraycolsep}{4pt}
\raise-11.1ex\hbox{\includegraphics[width=3.5cm]{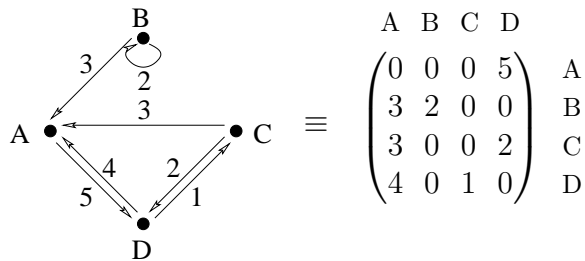}}\quad
  \raise-1.8ex\hbox{$\equiv$}\quad
    \begin{matrix}
      \hbox to 4.5em{\vphantom{\Big|}\footnotesize A\hfil B\hfil C\hfil D}
      \\
      \begin{pmatrix}
        0 & 0 & 0 & 5 \\
        3 & 2 & 0 & 0 \\
        3 & 0 & 0 & 2 \\
        4 & 0 & 1 & 0
      \end{pmatrix}
      &
      \begin{matrix}
        \mbox{\footnotesize A} \\
        \mbox{\footnotesize B} \\
        \mbox{\footnotesize C} \\
        \mbox{\footnotesize D}
      \end{matrix}
    \end{matrix} \nonumber
\end{equation}
\caption{A picture of a weighted directed network with $L=4$ nodes,
  labelled $A$, $B$, $C$ and $D$. The direction and weight of each
  link is shown. The corresponding adjacency matrix is also shown.} 
\label{fig:dwnet}
\end{center}
\end{figure}

We study a model  of a dynamically evolving, directed network of $L$
nodes with weighted links. Since the weights are integers we will
denote the weight associated with the link from node $k$ into node $l$
by an integer $n_{kl}\geq 0$ (instead of the usual $w_{kl}$). When
$n_{kl}=0$, this represents no link from $k$ to $l$. We depict such a
network with $L=4$ nodes in Figure \ref{fig:dwnet}.

The dynamics of the network are such that one unit of the weight
associated with the link pointing from node $k$ into node $l$ is
rewired to point into another randomly selected node $l'$ with a rate
$u_k(\underline{n}_l)$, where $\underline{n}_l \equiv n_{1l}, \ldots,
n_{Ll}$. In general, this rate is a function of the weights associated
with all of the links pointing into node $l$, and it depends on the
source node $k$: we will specialise to a rate of the form
\begin{equation} \label{rate}
u_k(\underline{n}_l) = u^s (n_{kl}) u^c(\sum_{k=1}^L n_{kl})\;,
\end{equation}
which depends on the weight of the link being rewired, through the
function $u^s(n_{kl})$, and the total in-strength of node $l$, through
the function $u^c(\sum_{k=1}^L n_{kl})$.  So for the example in Figure
\ref{fig:dwnet}, the rate at which the one unit of weight for the link
connecting node $C$ to node $A$ is rewired to another node is
$u_C(n_{AA}, n_{BA}, n_{CA}, n_{DA})=u^s(3) u^c(10)$. After the
rewiring, $n_{CA}=2$ and any of $n_{CB}$, $n_{CC}$ or $n_{CD}$ have
increased by one. These dynamics conserve the {\it out-strength}
$M_k=\sum_{l=1}^L n_{kl}$ which is the total weight of all links
pointing out of node $k$; the total weight of all links pointing into
node $l$, i.e., the {\it in-strength} $X_l=\sum_{k=1}^L n_{kl}$, is not
conserved.  For simplicity we will take $M_k = M$ for all $k$. Later,
in Section \ref{generalisations}, we will consider several
generalisations of the model and dynamics. In particular we consider
rewiring dynamics which depend on the node to which a link is being
rewired. This allows one to make a connection with a form of preferential
attachment in which links are preferentially rewired to nodes with
large in-strengths \cite{BBV04}. We proceed with the dynamics defined
above in order to make a clear connection to well-studied interacting
particle systems.  We stress that either alternative for the dynamics
yields the same behaviour and transitions.

\subsection{Mapping to a system of interacting particles} \label{mapping}
To make the connection with interacting particle systems, we use the
adjacency matrix, $A$, to encode the network: the value of element
$(k,l)$ of the matrix is $n_{kl}$, the weight of the link pointing out
of node $k$ into node $l$, as shown for example in Figure
\ref{fig:dwnet}. Note that here, because the network is directed,
$A\neq A^T$. The idea is to represent the adjacency matrix by a
lattice, where site $(k,l)$ of the lattice represents element $(k,l)$
of the adjacency matrix. The value of the element $(k,l)$, $n_{kl}$,
is then understood as the number of particles at site $(k,l)$ of the
lattice. The lattice is therefore two-dimensional, since it represents
the elements of a matrix, even though the network model may typically
be defined in a fully connected geometry (i.e.\ a link pointing into
one node is rewired to point into any other randomly selected node).

The network dynamics then causes the elements of the adjacency matrix
to change. Thus the corresponding particle configurations of the
lattice model also evolve according to dynamical rules, rules which
have their origin in the dynamics of the network model. 
For the network dynamics
described above the corresponding interacting particle system
is a generalisation of the much studied zero-range process \cite{EH05},
for which the steady state is exactly solvable.

\subsection{Model as a ZRP with $L$ species of particles}
The interacting particle system is defined on a square lattice (since
the adjacency matrix is a square matrix) with $L\times L$ sites. On
this lattice $n_{kl}$ labels the number of particles at the site
located at row $k=1,\ldots,L$, column $l=1,\ldots,L$.  There are $M$
particles in each row, a total of $LM$ in the system. Under the
network dynamics described in Section \ref{network model}, both $M$
and $L$ are conserved. Rewiring single units of the weight of a link
connecting node $k$ to node $l$ corresponds, in the interacting
particle system, to a particle in row $k$ and column $l$ hopping to
another randomly selected site \emph{within the same row} with a rate
$u_k(\underline{n}_l)$, given by (\ref{rate}).

\begin{figure}
\begin{center}
\includegraphics[scale=0.5]{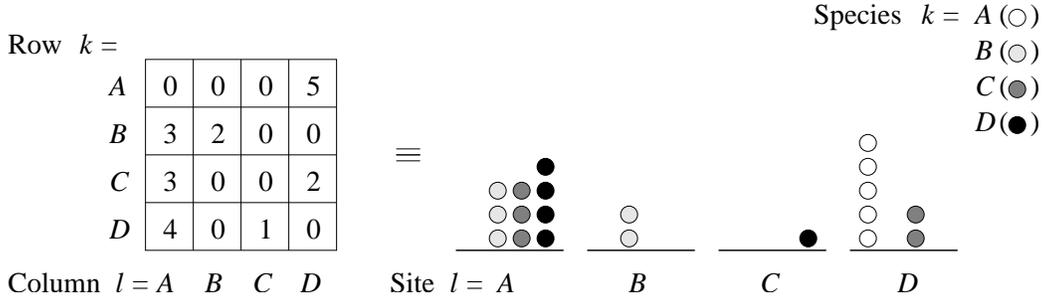}
\caption{A picture of the interacting particle systems which represent
the network depicted in Figure \ref{fig:dwnet}. On the left-hand side
is the two-dimensional lattice model with the occupation numbers
$n_{kl}$ entered in row $k$ and column $l$ of the lattice. On the
right-hand side is the corresponding multi-species ZRP (here there are
four species) --- note that there is no ordering amongst the particles
at a particular site.}
\label{fig:lats}
\end{center}
\end{figure}

We remark that these are the dynamics of a ZRP with $L$ species of particles,
labelled $k=1,\ldots,L$. Each site $l=1,\ldots,L$ of a one-dimensional
lattice contains $n_{kl}$ particles of species $k$. A particle of
species $k$ then hops with a rate $u_k(\underline{n}_l)$ to any other site on
the lattice, i.e.\ a rate which depends upon the number of particles of
each species at the departure site $l$. Thus we identify rows of the
two-dimensional lattice with different particle species (so the total
number of particles of each species is conserved) and the columns of
the two-dimensional lattice are identified with sites of a
one-dimensional lattice, as illustrated in Figure \ref{fig:lats}.
This mapping is useful because much is known about the exact steady states
of zero-range processes and many of its generalisations. The model
defined above can be solved exactly in the steady state: the analysis
which follows exploits this knowledge and one expects that many
variations of the network model will also be exactly solvable, those
which are can be inferred from the corresponding zero-range process.

Although the model we consider can be thought of as a ZRP with $L$
species of particles, we will proceed with the terminology of the rows
and columns of a two-dimensional lattice, since this retains closer
contact with the adjacency matrix of the original network model.

\section{Simulations} \label{simulations}

In this section we present results from simulations of the model that
show the two types of condensation: {\em site condensation}, where a
finite fraction of the particles in each row condenses onto a single
randomly  located
site of the row;  and {\em column condensation}, 
where a finite
fraction of all the particles in the system condenses into a single
column.  In the network context this corresponds to a finite
fraction of the weight from each node being contained in a single link
from that node; and a finite fraction of the weight pointing from all
nodes to a single node, respectively.  We also present simulations
which exhibit characteristics of both
types of condensation which are 
interesting from a networks perspective.

The model is simulated using a simple Monte Carlo algorithm.  At each
time step the following update procedure is followed:
\begin{enumerate}
\item A site of the lattice $(k,l)$ is chosen at random.
\item If the site is occupied then a particle will be removed from
this site with probability $u(\underline{n}_l) \delta t$, where
$u$ is the hop rate (\ref{rate}) and $\delta t$ is the time interval,
chosen such that the probability of a hop occurring is always less
than or equal to one.
\item If a particle was removed, then a second site $(k,m)$ (within 
the same row) is chosen at random and a particle is added to this site.
\end{enumerate}

All simulations were run on lattices of size $100 \times 100$, i.e.\
on networks with 100 nodes, and for $\mathcal{O}(10^{7})$
Monte Carlo Sweeps.  The first half of each run was used to relax to
the steady state and the second half to measure probability
distributions.  Distributions were measured for
\begin{itemize}
\item the number of particles at a site (link weight)
\item  the number of particles in a column
(node in-strength)
\item  the number of occupied sites in a column and
row (in- and out-degree respectively).
\end{itemize}
Typical configurations of the lattice were also output.

In order to compare with theory (see Section \ref{theory}), the
following hop rates were chosen to show the two condensation types.
Recall that we consider a hop rate for particles from site $(k,l)$ of
the form (\ref{rate}). For site condensation we make the choice 
\begin{equation} \label{site rate}
u^s(n) = (1+b^s/n)  \qquad \textrm{and} \qquad u^c(n) = 1\,,
\end{equation}
with $b^s = 4$ and for $n>0$. For column condensation 
we make the choice 
\begin{eqnarray} \label{col rate}
u^c(n) = \left\{ \begin{array}{ll}
1 + b^c & \textrm{for $n \leq L$}\,, \\ 
1 + b^c L/n & \textrm{for $n > L$}\,,
\end{array} \right. \qquad \textrm{and} \qquad u^s(n) = 1\,,
\end{eqnarray}
with $b^c = 1.05$.

\subsection{Site Condensation} \label{site condensation}
\begin{figure}
\begin{center}
\includegraphics[width=12cm]{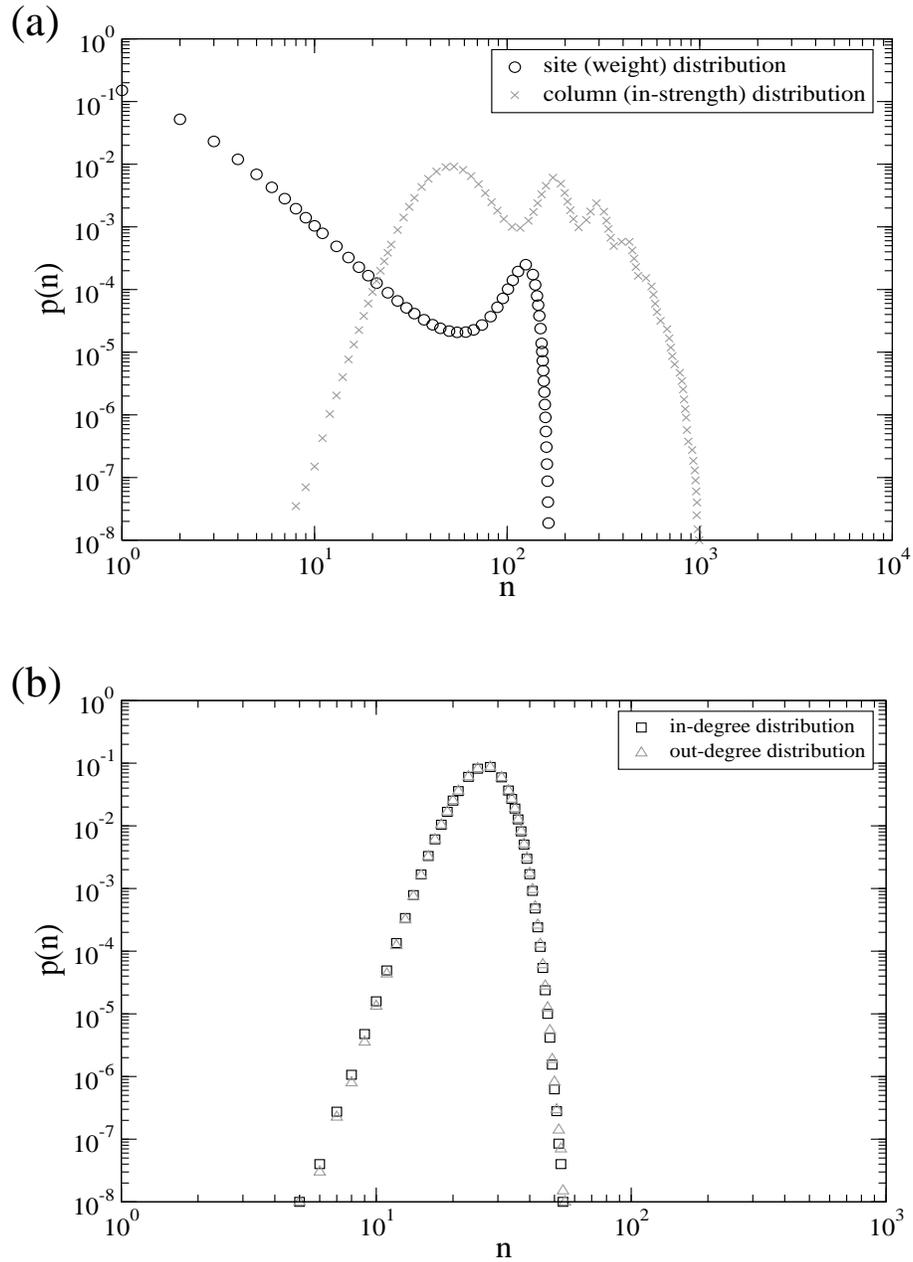}
\caption{Distributions from simulation of the model with 
the hop-rate chosen to display site condensation.  
(a) Site and column occupation distributions, corresponding 
to the distributions of weight among links and in-strength 
among nodes in the network context respectively. 
(b) In- and out-degree distributions.}
\label{simfigsc}
\end{center}
\end{figure}

\begin{figure}
\begin{center}
\includegraphics[width=12cm]{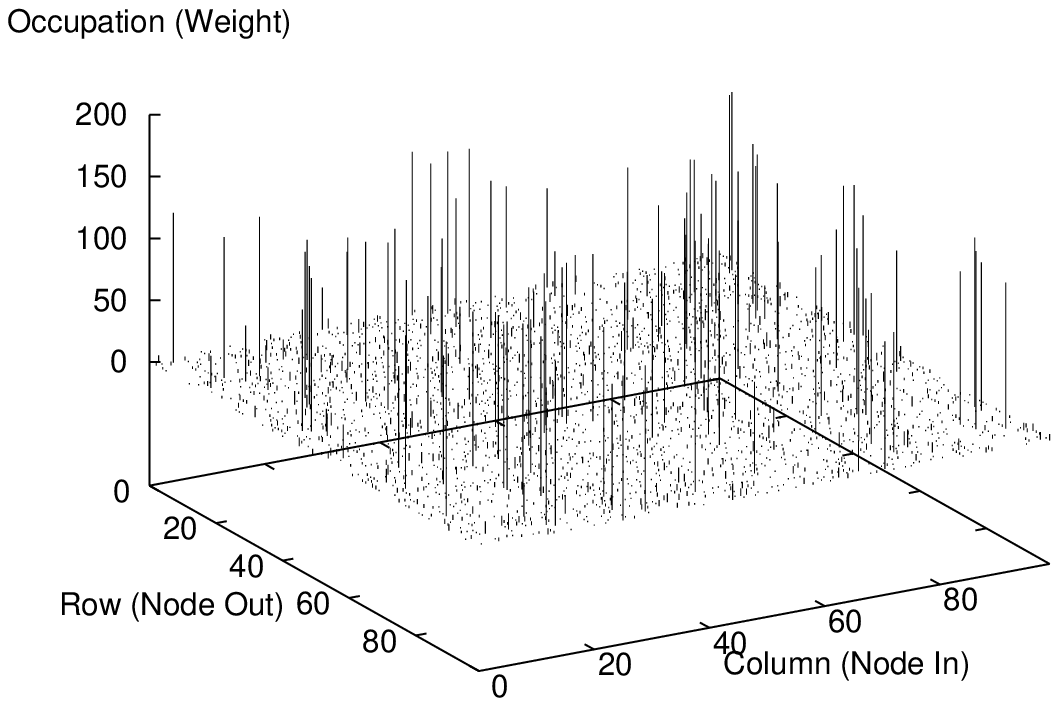}
\caption{Typical configuration of the system, output from simulation 
of the model with the hop-rate chosen to display site condensation.}
\label{simfigscconf}
\end{center}
\end{figure}

The simulations for the site condensation case were run with $175$
particles in each row starting from a random initial configuration. The
site condensation can be seen clearly from the site distribution
(black circles, Figure~\ref{simfigsc} (a)).  The peak at around
$n=150$ represents the site condensates, it has an area of order $1/L$
and there are $L^2$ sites indicating that $L$ sites have occupations
of around $150$ particles.  As the number of particles in each row is
fixed at $175$, there can be at most one such site in a row and an
occupation of this size represents an appreciable fraction of the
total number of particles available to a site, hence implying a
condensation in the thermodynamic limit.  Thus we have a single
condensed site in each row.  Note also that the site distribution has
an apparently power-law background before the condensate peak, this is
reminiscent of the behaviour of a single-species ZRP \cite{EH05},
indeed as the column attraction has effectively been switched off, the
system is a collection of single-species ZRPs.  The column
distribution (grey crosses, Figure~\ref{simfigsc} (a)) shows that, in
the absence of any coupling between the rows, the condensed sites are
randomly distributed among the columns --- as one would expect.  The
first peak in the distribution at around $n=50$ shows that some
columns do not contain any condensed sites, just those from the
power-law background.  The second, third, $\ldots$ peaks correspond to
columns with one, two, $\ldots$ condensed sites in them, plus a
background from the rest of the sites.  This is exactly what one would
expect if the condensed sites were randomly distributed in the columns
and a random sampling of the site distribution to create a column
distribution agrees closely with the simulation.  The typical
configuration (Figure~\ref{simfigscconf}) bears this out, condensates
can be seen on individual sites, but no column ordering can be
discerned.  Although not shown here, systems with weak column
attraction showed similar behaviour, i.e. as though no column
attraction was present.

The degree distributions for the site condensation case
(Figure~\ref{simfigsc} (b)) appear to be binomial in nature: a
binomial distribution constructed from the measured probability of a
site having zero occupation matches very closely the data shown.  Thus
in the network context, for the site condensation we have connectivity
similar to that of a random graph (which also has a binomial 
connectivity distribution). In addition to this we have
a condensed link weight from each site with a power-law background,
and an in-strength which is a random sum of link weights.

\subsection{Column Condensation}  \label{column condensation}
\begin{figure}
\begin{center}
\includegraphics[width=12cm]{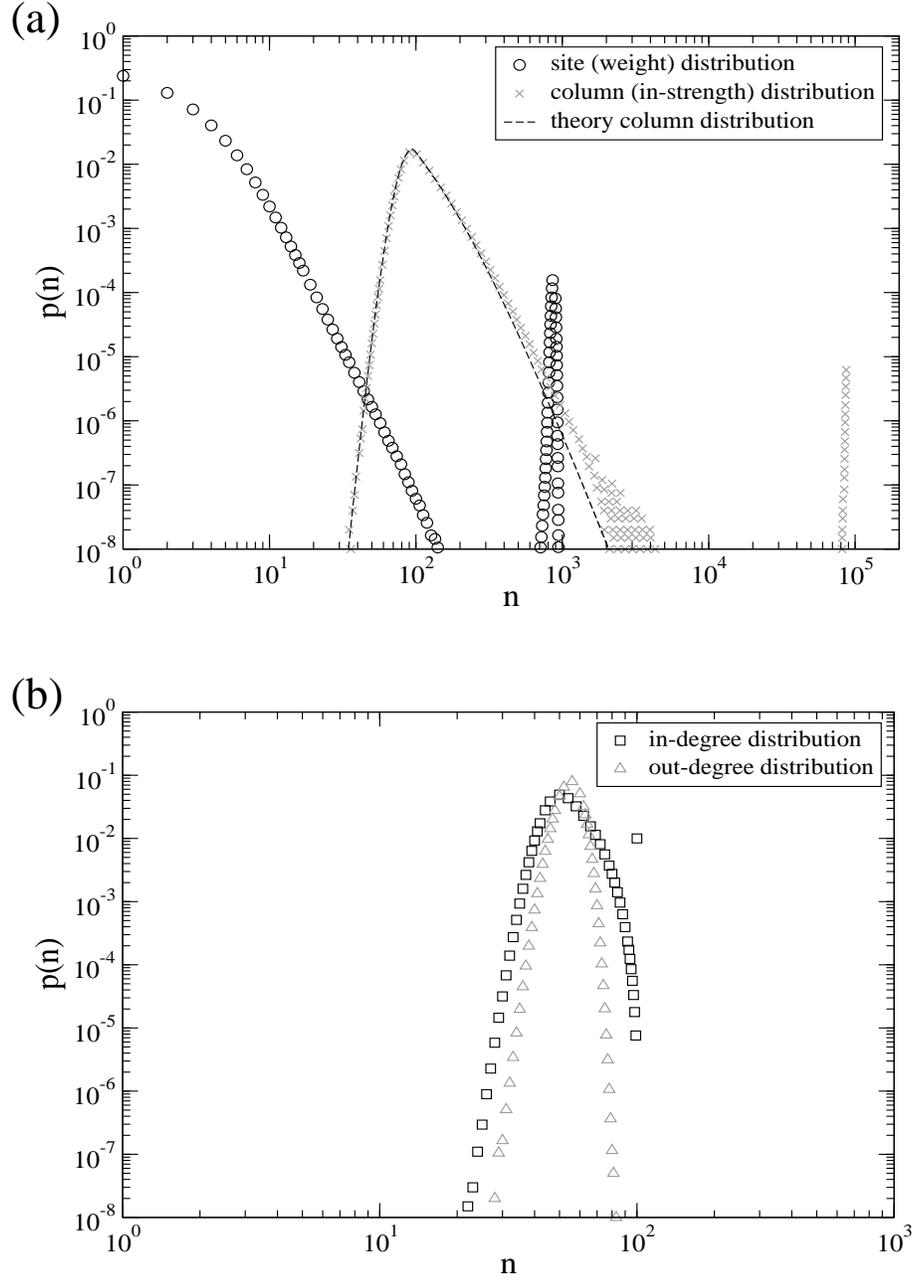}
\caption{Distributions from simulation of the model with the
hop-rate chosen to display column condensation.
(a) Site and column occupation distributions, corresponding to 
the distribution of weight among links and in-strength among
nodes in the network context respectively.
(b) In- and out-degree distributions.}
\label{simfigcc}
\end{center}
\end{figure}

\begin{figure}
\begin{center}
\includegraphics[width=12cm]{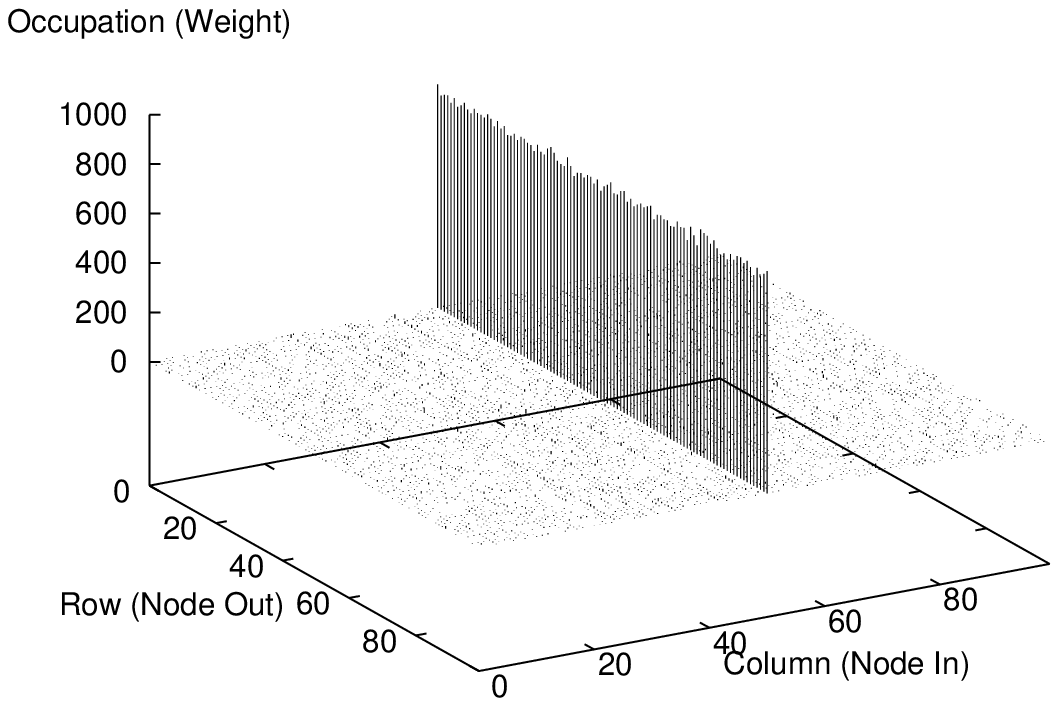}
\caption{Typical configuration of the system output from simulation of the 
model with the hop-rate chosen to display column condensation.}
\label{simfigccconf}
\end{center}
\end{figure}

The simulations for the column condensation case were run with $1000$ 
particles in each row. The initial condition was taken with all
particles on the same site.  
The column condensation can be seen clearly from the column
distribution (grey crosses, Figure~\ref{simfigcc} (a)).  The peak at
around $n=85000$ corresponds to the condensate, it has an area of
order $1/L$ and there are $L$ columns, indicating that a single column
has an occupation of around $85000$, an appreciable fraction of the
total number of particles in the system $L\times M = 100 \times 1000$.
The other, lower peak at around $n=100$ represents the columns which do
not contain a condensate.  The site distribution (black circles,
Figure~\ref{simfigcc} (b)) also shows condensation onto sites in the
peak at around $n=850$.  If the column interaction were to be switched
off leaving $L$ uncoupled single-species ZRPs, it is known that the
chosen $u^s(n)$ ($= \mathrm{constant}$) would not give condensation in 
this fully connected homogeneous system, see for example \cite{EH05}.  
Thus in this case it is
the column interaction that has induced a site condensation.  
Note that both the column and site distributions have apparent 
power-law pieces, with equal or close exponents.  The theory 
presented in Section~\ref{theory} allows one to construct the 
critical column distribution at the transition point --- shown as 
the dotted line in Figure~\ref{simfigcc}~(a).  This fits the 
non-condensate background part of the measured column distribution 
very well. The typical configuration of the system, 
Figure~\ref{simfigccconf},
shows the column condensate comprises site condensates all in the
same column.

The out-degree distribution for the column condensation case (grey
triangles, Figure~\ref{simfigcc}~(b)) is similar to the site
condensation case, being distributed according to a binomial
distribution.  However, the in-degree distribution
(Figure~\ref{simfigcc}~(b), black squares) is significantly different,
in particular it has a broader tail and also a single site which is
connected to all others, as shown by the lone data point at degree $L$
with probability $1/L$ ($L=100$ in this case). Thus in the network
context the out-connectivity resembles that of a
random graph, but
the in-degree distribution displays a broader tail and there is a
single node to which all others point --- the node corresponding to
the condensed column. Furthermore, all links pointing to this node
hold a finite fraction of the weight available and the node holds a
finite fraction of the total in-strength available to it.
Broadly-tailed degree distributions are often observed in real
networks, weighted and unweighted alike.  While the tail observed 
in these simulations is not as broad as many observed, it is at least 
broader than that of the equivalent random graph.  Also apart from the
condensed pieces the in-strength and weight distributions display
power-law tails, again something that has been observed in many real
networks.  Thus, as with equilibrium networks \cite{DM03,DMS03},
certain distributions may take a power-law form at a critical point.
Power-law strength and weight distributions and non-power-law degree
distributions have been observed for a traffic network in
\cite{MBCV05}.

As mentioned at the beginning of this subsection, we employed a fully
condensed initial condition. The reason is that, for a random initial
configuration, the dynamics slows down as the system reaches a state
in which several columns contain a large number of particles; the
single column condensate is attained only on prohibitively long
timescales.

\subsection{Further behaviour}
In the preceding sections the somewhat extreme cases of the absence of
site or column attraction in the presence of the other ($u^s$ or $u^c$
set equal to one) were considered.  This was to allow for a direct
comparison with theory.  Cases where some forms of non-trivial
attraction are present in both $u^s$ and $u^c$ are difficult to
compare directly with theory.  However, such systems do display
interesting behaviour and so in this section we present some numerical
data from simulations of such a system.

Simulations were run on a system with $175$ particles in each row and
the hop-rate given by: $u^c(n) = 1 + b^c/L$ for $n\leq L$, $u^c(n) = 1
+ b^c/n$ for $n>L$ and $u^s(n) = 1 + b^s/n$, with $b^c = 16$ and $b^s
= 2.5$.  If $u^c$ were set equal to one, then no condensation would
occur at this particle number \cite{EH05}.  If $u^s$ were set equal to
one, then no condensation would take place at
any finite particle number (see Section~\ref{theory}). Results from
simulations show that this system, with both site and column
attraction present in forms that separately show no condensation
behaviour, does show something that could be interpreted as a
condensation-like phenomenon. The following results were obtained from
a random initial configuration.

\begin{figure}
\begin{center}
\includegraphics[width=12cm]{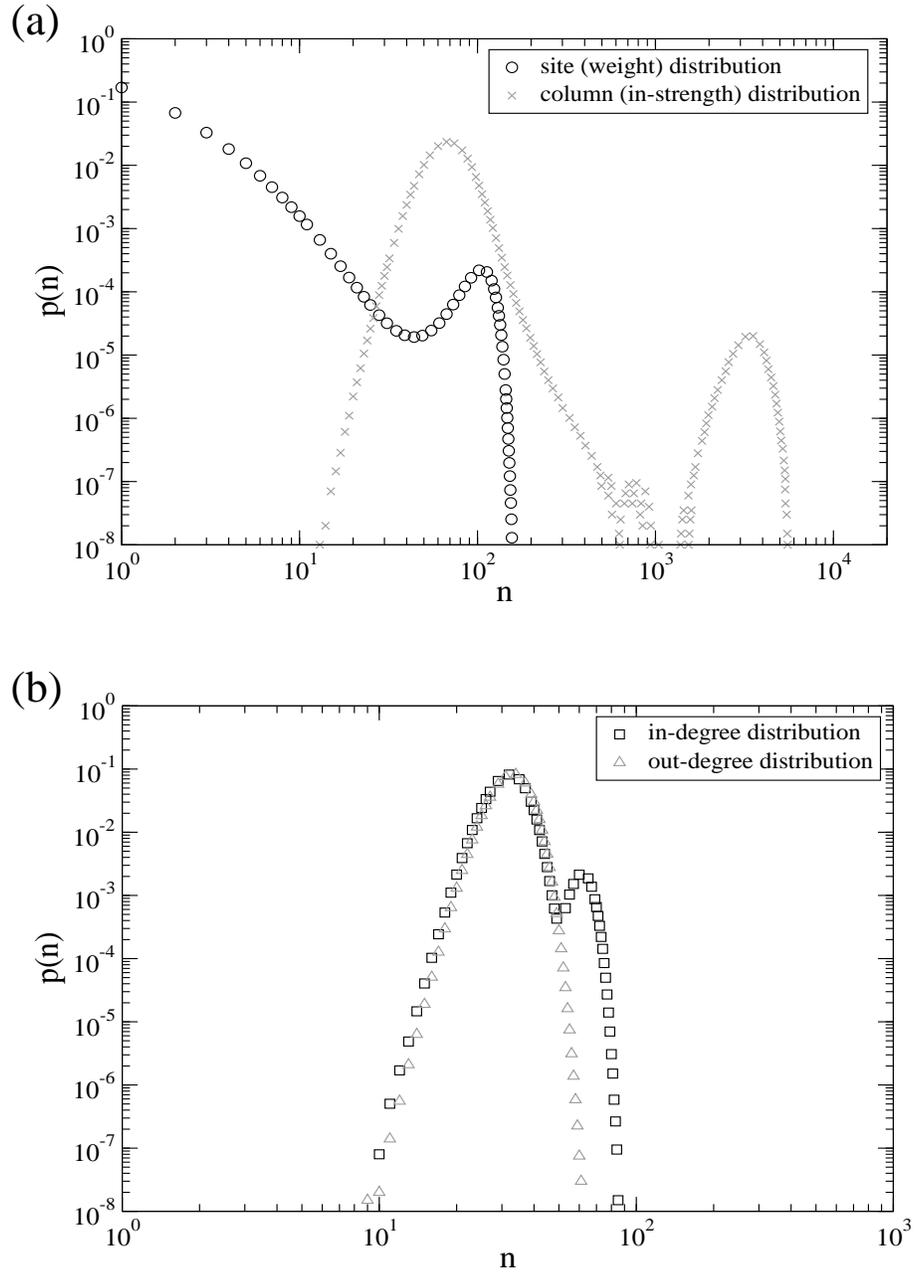}
\caption{Distributions from simulation of the model with 
the hop-rate chosen such that without the site or 
column attraction elements no condensation would be 
present.
(a) Site and column occupation distributions, corresponding 
to the distributions of weight among links and in-strength 
among nodes in the network context respectively. 
(b) In- and out-degree distributions.}
\label{simfigo}
\end{center}
\end{figure}

\begin{figure}
\begin{center}
\includegraphics[width=12cm]{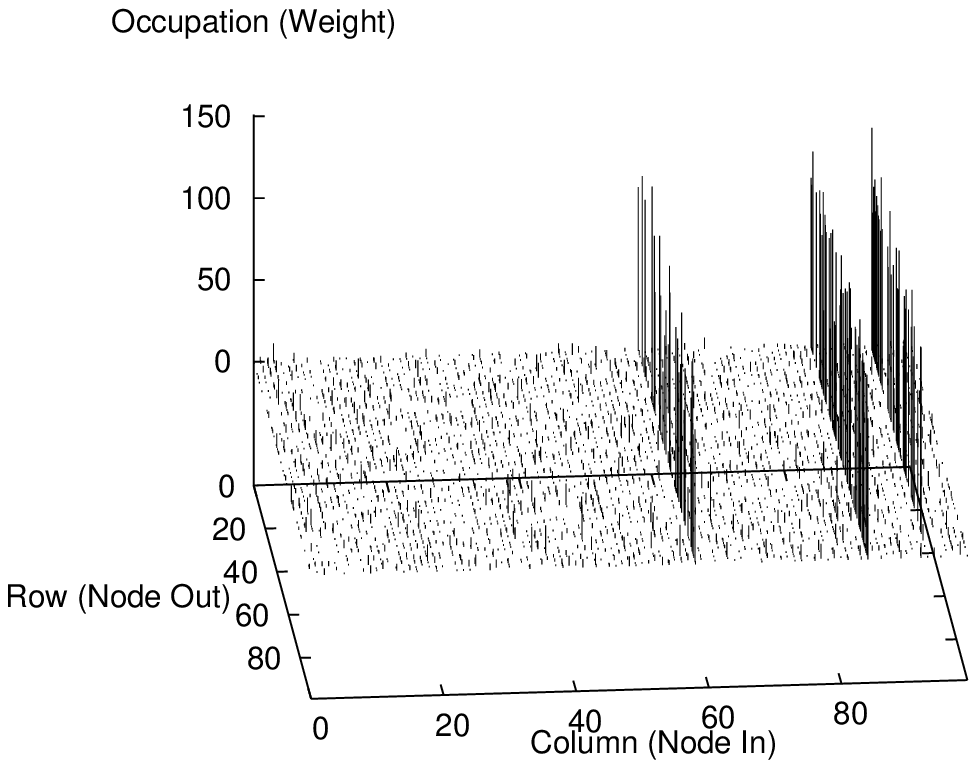}
\caption{Typical configuration of the system output from simulation of the 
model with the hop-rate chosen such that without the site 
or column attraction elements no condensation would be present.}
\label{simfigoconf}
\end{center}
\end{figure}

In Figure~\ref{simfigo} (a), it can be seen that the site occupation
distribution (black circles) has a decaying part and a peak at around
$n=150$ and is generally reminiscent of a condensed system.  The
column distribution (Figure~\ref{simfigo} (a) grey crosses) also shows
a large $n$ peak, at around $n=3500$, although it is much broader than
one usually associated with a clear condensation.  The typical
configuration data in Figure~\ref{simfigoconf} indicate that the high
$n$ bump of the
column distribution represents more than one highly occupied
column, each composed of many site condensates.  Without a direct
comparison with theory available, it is difficult to
interpret this broad peak: it could be a finite-size effect that will
not be seen in larger systems, but on the other hand it could be a
true condensation that is being adversely affected by the finite size
of the system.  In either case it is interesting from a network point
of view as real-world networks are often of finite size.

The degree distributions of this system are also interesting.  They both 
have a binomial form for low degree, but the in-degree distribution departs 
from this at high degree --- it has a small secondary peak implying that 
several sites are highly connected, but a completely connected site to 
which all others point, as arises in the column condensation case, is absent.
Thus in the network context we have a network with connectivity much 
like that of a random graph, except for the existence of several 
well-connected `hub' nodes which would not be present in the random case. 
These hub nodes also have many high weight links pointing to them, 
giving them a large in-strength.

For an initial configuration with all particles located on a single
site, we obtain distributions of the same form as above and configurations
typically contain more than one highly occupied column.

\section{Theory}\label{theory}
In this section we present the exact steady state of the model and
exploit this solution in order to understand theoretically the
condensation transitions observed in simulations. 

\subsection{Steady state}

First, we give the steady state, the proof of which is given in an
appendix. The steady state probabilities
$P(\{\underline{n}_{l}\})$, for finding the system in a configuration
$\{\underline{n}_{l}\}$, are given by the simple form
\begin{equation} \label{P(C)}
P(\{\underline{n}_{l}\}) = Z_{L,M}^{-1} \prod_{l=1}^L f(\underline{n}_l)\;,
\end{equation}
i.e.\ a product over columns. Here, the functions $f(\underline{n}_l)$
depend on a single column (i.e.\ the in-links of node $l$), and for
the rates (\ref{rate}), they are given by
\begin{equation} \label{f(n)}
f(\underline{n}_l) = f^c(X_l) \prod_{k=1}^L f^s(n_{kl}) \;,
\end{equation}
where we are using $X_l = \sum_{k=1}^L n_{kl}$ to represent the total
number of particles in a column (the node in-strength), and where
\begin{equation} \label{f^a(x)}
f^a(x) = \prod_{i=1}^x u^a(i)^{-1}\;,
\end{equation}
for $a=s,c$, where $f^a(0)=1$.
The normalisation $Z_{L,M}$ in (\ref{P(C)}) is given by
\begin{equation} \label{Z_{L,M}}
Z_{L,M} = \sum_{\{n_{kl}\}} \,\prod_{l=1}^L \left[ f^c(X_l) \prod_{k=1}^L
f^s(n_{kl})  \right] \,\prod_{k=1}^L\delta(\sum_{l=1}^L
n_{kl} - M)\;. 
\end{equation}
The $\delta$-function in this normalisation enforces particle
conservation within each row (which is the conservation of node
out-strength in the network): it is this $\delta$-function that induces
correlations between different columns in the steady state.
 
\subsection{Condensation theory}
Now, we exploit the exact steady state in order to understand
theoretically the condensation transitions observed in simulations.
Ideally, one would wish to demonstrate condensation in the site- and
column-distributions of particles, $p^s(n)$ and $p^c(n)$. The
expressions for these distributions involve the normalisation
$Z_{L,M}$ given by (\ref{Z_{L,M}}), which can be thought of as a
canonical partition function since $M$ is fixed. However, as we now
outline, it turns out to be simplest to work within the grand
canonical ensemble in which the particle number is allowed to
fluctuate.

\subsubsection{Grand canonical ensemble}
We introduce the grand canonical partition function in the usual way,
by defining fugacities $\{z_k\}$ so that we can replace the canonical
partition function (\ref{Z_{L,M}}) by 
\begin{equation} \label{GCZ1}
\mathcal{Z}_L = \sum_{\{n_{kl}=0\}}^\infty \prod_{k=1}^L
z_k^{\sum_l n_{kl}} \prod_{l=1}^L\,\left[
 f^c(X_l) \prod_{k=1}^L f^s(n_{kl}) \right] \;,
\end{equation} 
where the fugacities are chosen to ensure that, on average, each row
contains the proper number of particles $M_k$:
\begin{equation} \label{fugacity}
z_k \frac{\partial \, {\rm ln} \mathcal{Z}_L}{\partial z_k} =
\overline{\sum_l n_{kl}} = M_k\;.
\end{equation}
Here, the overbar indicates an average taken in the grand canonical
ensemble. Since we are interested in the case where $M_k = M$ for all
values of $k$, the lhs of this equation must be independent of $k$
therefore $z_k = z$ for all values of $k$; in this case, the fugacity $z$
is chosen to determine the {\it total} number of particles in the
system. After a little rearrangement, (\ref{GCZ1}) can be written
\begin{eqnarray} \label{GCZ2}
\mathcal{Z}_L &=&
 \prod_{l=1}^L\,\left[ \left( \prod_{k=1}^L
 \sum_{n_{kl}} \right) f^c(X_l) \prod_{k=1}^L
 f^s(n_{kl})\, z^{n_{kl}} \right] \\
&=&
 \left[ 
 \sum_{\{ n_{k}\}}  f^c(X) \prod_{k=1}^L
 f^s(n_{k})\, z^{n_{k}} \right]^L \;,
\end{eqnarray} 
where in going from the first to the second line
we use the fact
that the $l$ subscript on the $n_{kl}$
plays no role --- each column makes the same contribution to
$\mathcal{Z}_L$ --- and the square bracket can simply be raised to the
power of $L$. Now, $X = \sum_k n_k$. By taking
derivatives of $\mathcal{Z}_L$ with respect to $z$, one finds that the
fugacity must be chosen to satisfy
\begin{equation} \label{saddle}
M = z \frac{\partial\,{\rm ln} F(z)}{\partial z}\;,
\end{equation}
where we have defined the function
\begin{equation} \label{F(z)}
F(z) = \sum_{\{n_k\}} f^c(X) \prod_{k=1}^L
 f^s(n_{k})\, z^{n_{k}} \;.
\end{equation}

The equation (\ref{saddle}) is the key result of this
subsection. Condensation typically arises when one is unable to
satisfy (\ref{saddle}) above some critical density; under these
circumstances, the grand canonical ensemble is no longer valid and the
change of the partition function from the grand canonical form below
the critical density to some other form above it signals the phase
transition. We now discuss conditions under which (\ref{saddle}) can
be satisfied in order to illustrate how one can theoretically
understand and predict the condensed phases observed in simulations.

\subsubsection{Condensation} \label{Condensation}
It is straightforward to show that the function $F(z)$ is a smoothly
increasing function of its argument $z$. Let us assume that we have
chosen hop rates such that the sums which determine $F(z)$ have a
radius of convergence $\alpha$. The convergence properties of $F(z)$
and its derivative determine whether or not the system undergoes
condensation. These convergence properties are determined by the form
of the function $f(\underline n)$ which in turn provides the conditions
on the rewiring rates under which the system condenses. We now present
two very simple examples to illustrate the two types of condensation.

First, we consider conditions to observe site condensation. We
take $f^c(x) = 1$, which is the case when $u^c(x) = 1$ for all $x$, hence
\begin{equation}
F(z) = \left[ \sum_{n=0}^\infty f^s(n) z^n \right]^L\;.
\end{equation}
Hence, from (\ref{saddle}), the fugacity is determined by
\begin{equation}
\rho = \frac{\sum_{n=0}^\infty n f^s(n) z^n}{\sum_{n=0}^\infty f^s(n)
  z^n}\;,
\end{equation}
where $\rho= M/L$ is the density of particles in each row. When both
the numerator and denominator approach finite values as $z$ approaches
the radius of convergence $\alpha$ there exists a finite critical
density $\rho_c$ above which (\ref{saddle}) cannot be satisfied. This
signals condensation: for $\rho > \rho_c$, each row contains $(\rho -
\rho_c)L$ `excess' particles which condense in a single, randomly
located site within each row. In order that $\rho_c$ is finite, one
requires that $f^s(n)$ decays asymptotically like $f^s(n) \sim
\alpha^{-n} n^{-b}$ for large $n$ where $b>2$. This in turn implies
that the hop rates $u^s(n)$ must decay asymptotically as
\begin{equation} \label{u^s(n)}
u^s(n) \sim \alpha(1+b/n) \;,
\end{equation}
again, where $b>2$, if one is to observe site condensation in each row
at finite density. In the network model, this implies that the rate of
rewiring a node must decay more slowly than $2/n$, where $n$ is the
weight of the link pointing out of the node being rewired. These
considerations guide the choice of hop rates (\ref{site rate})
discussed in Section \ref{site condensation}.

Now, we consider conditions to observe column condensation. To deal
with the sum over particle configurations in $F(z)$, it turns out to
be convenient to sum over the variable $X$ and introduce a
$\delta$-function to ensure that $X$ represents the number of
particles in a column. Thus $F(z)$ may be written in the form
\begin{equation}
F(z) = \sum_{\{n_{k}\}} \sum_{X=0}^\infty f^c(X) \,
 \delta(\sum_{k=1}^L n_{k} - X) \prod_{k=1}^L f^s(n_{k})
 z^{n_{k}}\;,
\end{equation} 
where the $k$ subscripts refer to the rows within a single column. To
proceed, we take $f^s(n) = 1$, which is the case when $u^s(n) =
1$ for all $n$. With this choice, the sum over $\{n_k\}$
is straightforward to perform --- the $\delta$-function constraint
supplies a combinatorial factor associated with the number of ways of
adding $L$ integers to get $X$ --- and one finds
\begin{equation}
F(z) = \sum_X {L{-}1{+}X \choose L{-}1}\,f^c(X)z^X\;.
\label{F(z)X}
\end{equation}
In this case, the rhs of (\ref{saddle}) represents the average number
of particles in a column, $\overline X$. The asymptotics of $f^c(X)$
determine the convergence properties of $F(z)$ (\ref{F(z)X}). 
In order to deduce the
required asymptotics, we expand the binomial factor for $X \gg L$,
using the approximation 
\begin{equation}
{L{-}1{+}X \choose L{-}1} \approx \frac{X^{L-1}}{(L-1)!}\,,
\end{equation}
for large $X$. To observe column condensation, the rhs of
(\ref{saddle}) must converge to a value $X_c=\mathcal{O}(L)$ as $z \to
\alpha$. This is the case when $f^c(X)$ decays asymptotically like
$f^s(n) \sim \alpha^{-X} X^{-\lambda}$ with $\lambda > 1+L$. When
these conditions are satisfied, the system condenses above a critical
density $\rho_c = X_c/L$ in such a way that every site in a single,
randomly located column typically contains $(X-X_c)$ particles. A
hop rate which yields $f^c(X)$ with the required asymptotics is
\begin{equation} \label{u^c(n)}
u^c(X) \sim \alpha(1+c L/X) \;,
\end{equation}
where $c>1$, which motivates the choice of hop rates (\ref{col rate})
discussed in Section \ref{column condensation}.

Thus these two simple examples, $f^c(x)=1$ and $f^s(n)=1$, illustrate
site and column condensation respectively.  We would like to stress
that in both site and column condensation, while only the asymptotics
determine whether or not the system condenses, the actual value of the
critical density depends on the details of the form of the rewiring
rates for all values of their arguments, not just the asymptotics.

We now indicate how one can use the grand canonical ensemble to
predict the background particle distribution in the condensed
phase. We consider the case $f^s(n)=1$, for which $p^c(X)$, the
probability that a column contains exactly $X$ particles, is given by
\begin{equation} \label{p(X)}
p^c(X) = \frac{{L{-}1{+}X \choose L{-}1}\,f^c(X)z^X}{F(z)}\;.
\end{equation}
This equation holds throughout the low density phase. In the condensed
phase, the grand canonical ensemble breaks down, however (\ref{p(X)})
with $z=\alpha$ correctly reproduces the form of the background
distribution of column occupation numbers. For the rates (\ref{col
rate}), (\ref{p(X)}) cannot be written in a convenient form at $z=1$,
however it can easily be evaluated numerically for a given value of
$b^c$. The result of such a computation, with $b^c = 1.05$, is
compared with simulation in Figure~\ref{simfigcc}~(a).

\section{Generalisations}
 \label{generalisations}

In this section we consider some generalisations of the network model
which, in the steady state, are still given by a factorised form. The
aim is to illustrate that the class of network models with factorised
steady states extends to a wide variety of rewiring dynamics and so,
by making the connection between network dynamics and interacting
particle systems, we are rewarded with a versatile approach to the
analysis of network properties.

\subsection{General dependence of the rewiring rates on departure and
  destination nodes} 

We begin by considering three generalisations of the rewiring rates: 
\begin{enumerate}
\item General dependence of the rewiring rates on
the weights of links rather than the specialised form (\ref{rate}).

\item Rewiring rates which depend on the
weights of the links pointing into both the node from which the link
is being removed, and the weights of the links pointing into the 
node into which the link is being rewired --- this allows one to
consider preferential attachment.

\item Heterogeneity in the rewiring dynamics such that one can
  consider rates which differ depending on the source, departure and
  destination nodes of the link.

\end{enumerate}
With these generalisations, one unit of the weight associated with the
link pointing from node $k$ into node $l$ is rewired to point into
another randomly selected node $l'$ with a rate
$u_{kl}(\underline{n}_l) \, t_{kl'}(\underline{n}_{l'})$. The
heterogeneity described in the above point 3 enters through the $k$,
$l$, and $l'$ superscripts on $u$ and $t$. The corresponding
interacting particle system has been discussed in the literature in
the context of `Urn' models (discussed in \cite{GL02}) and
`Misanthrope' processes (discussed in \cite{EH05}) in a fully
connected geometry, in which particle hop rates depend on the numbers
of particles at both the site of departure and the destination site,
but now defined with $L$ species of particles.

In the steady state, this model can still be expressed in the
factorised form 
\begin{equation} \label{P(C)gen}
P(\{\underline{n}_{l}\}) = Z_{L,M}^{-1} \prod_{l=1}^L f_l(\underline{n}_l)\;,
\end{equation}
(though note $f_l(\underline{n}_l)$ now carries a sub-script $l$)
provided the hop rates satisfy the constraint
\begin{eqnarray} \label{constraint}
 \frac{u_{kl}(\underline{n}_{l}; n_{k'l}{-}1) \,
 t_{kl}(\underline{n}_{l}; n_{kl}{-}1)}
{u_{kl}(\underline{n}_{l}) \,
 t_{kl}(\underline{n}_{l}; n_{kl}{-}1, n_{k'l}{-}1)} = 
 \frac{u_{k'l}(\underline{n}_{l}; n_{kl}{-}1) \,
 t_{k'l}(\underline{n}_{l}; n_{k'l}{-}1)}
{u_{k'l}(\underline{n}_{l}) \,
 t_{k'l}(\underline{n}_{l}; n_{kl}{-}1, n_{k'l}{-}1)} \;,
\end{eqnarray}
for all pairs of rows $k$ and $k'$. We have introduced the notation
$\underline{n}_{l}; n_{kl}{-}1$ which represents the configuration
$\underline{n}_l$ but with the term $n_{kl}$ replaced by
$n_{kl}{-}1$. When the hop rates satisfy this constraint, the
functions $f_l(\underline{n}_l)$ can be written
\begin{equation} \label{f(n)gen}
f_l(\underline{n}_l) = \prod_{k=1}^L \left[ \prod_{i=1}^{n_{kl}}
  \frac{t_{kl}(0,\ldots,0,i{-}1,n_{k+1\,l},\ldots,n_{Ll})}
{u_{kl}(0,\ldots,0,i,n_{k+1\,l},\ldots,n_{Ll})} \right]\;,
\end{equation}
having chosen $f(0,\ldots,0) = 1$. This can be written in a number of
different forms due to (\ref{constraint}) --- the symmetry in
$f_l(\underline{n}_l)$ is obscured within this constraint. A proof of
this steady state is presented in Appendix B.

\subsubsection{Preferential attachment}
To illustrate how one might exploit this steady state, we consider a
network model with preferential attachment rewiring dynamics. Thus we
choose  $u_{kl}(\underline{n}_{l}) = 1$ for all $k$ and $l$, and 
\begin{equation}
 t_{kl'}(\underline{n}_{l'}) = t^s(n_{kl'}) t^c (\sum_{k=1}^L n_{kl'})\;,
\end{equation}
which defines the rate a link pointing from node $k$ into node $l$ is
rewired to point into node $l'$. This rate therefore is a function of
the weight of the link pointing from the source node $k$ into the
destination node $l'$ before the rewiring event, and a function of the
total in-strength of the destination node. It satisfies the
constraint (\ref{constraint}) and one finds that the functions
$f(\underline{n}_l)$ are given by (\ref{f(n)}), where
\begin{equation}
f^a(x) = \prod_{i=1}^x t^a(i-1)\;,
\end{equation}
for $a=s,c$, where $f^a(0)=1$. In order that this model will exhibit
the behaviour observed in simulations, one can follow the analysis of
Section \ref{Condensation} to deduce that if the hop rates decay
asymptotically like
\begin{equation}
t^s(n) \sim 1-b/n\;,
\end{equation}
with $b>2$, then one observes site condensation above some finite
critical system density, and if
\begin{eqnarray}
t^c(X) \sim 1-\lambda/X \;,
\end{eqnarray}
for $X \gg L$, with $\lambda>L+1$, then one observes row condensation
above some finite critical system density.

\subsection{Further generalisations}
There are a number of further ways one can choose to redefine the
model and still obtain a factorised steady state. One way is to
consider geometries other than the fully connected one, such as
rewiring dynamics in which the weight being rewired from node $l$ is
always rewired to a `neighbouring' node $l+1$. All the models
considered so far also have factorised steady states in this geometry
although there may exist extra constraints in certain
cases. Generalisations to more complicated geometries are also
possible, and have been discussed in the context of the ZRP in
\cite{Evans00}.

We can also relax the property of integer weights and consider the
case where the weights are continuous variables and an arbitrary
amount of the weight is allowed to be rewired.  The approach is
similar to the way one generalises the zero-range process to
continuous masses \cite{EMZ}.

It is straightforward to allow each node to have a different total
out-strength, but more complicated generalisations whereby these
out-strengths are not conserved may also be possible. Non-conservation
in the context of the ZRP has been discussed in \cite{AELM05, EH05}. Another
straightforward modification of the model is to prohibit links which
point into the same node they point out of, i.e., restrict to
$n_{kk}=0$.  A further possibility is to allow negative weights, as
well as positive, which corresponds to a generalisation of the
so-called `Bricklayers' model' \cite{B03} --- positive and negative
weights have been considered in a rewiring social network in
\cite{HJDXW05}.

A point about the factorised steady states we would like to emphasise
is that one is free to choose any form for $f(\underline{n}_l)$ and
then infer the rewiring rates from a recursion such as
(\ref{recursion}). 

\section{Conclusion} \label{conclusion}

In summary we have introduced a dynamical model of a weighted, directed
network wherein the out-strength of each node is conserved but the
in-strength is not.  The rewiring dynamics of this model can be mapped
onto a multi-species zero-range process for which the steady state is
exactly solvable.

Through numerical simulations and theoretical analysis we have
identified two condensation transitions in the steady state.  In the
network context they correspond to a disparity transition where for
each node a single link contains a finite fraction of the
out-strength, and the more familiar condensation transition where the
in-strength of a single node captures a finite fraction of the
out-strengths for all nodes in the system.  These transitions
demonstrate some of the varied behaviour that is possible in weighted
directed networks.

Within the multi-species ZRP picture these transitions correspond to,
in the first case, condensation of all species at independent sites
or, in the second case, a collective condensation of all species onto
the same site.
From our knowledge of the multi-species ZRP we are able to identify
a number of generalisations of the model and dynamics which preserve the 
exactly solvable steady state. For example
these correspond in the network context to continuous weights,
preferential attachment dynamics and node fitness.
It would be interesting to explore further  the 
different possible behaviours afforded by these and other
generalisations.

\subsection*{Acknowledgements}
This work was supported by the Scottish Universities Physics Alliance
(SUPA) and EPSRC programme grant GR/S10377/01.  AGA thanks the
Carnegie Trust for the Universities of Scotland for financial support.

\appendix

\section*{Appendices}

\setcounter{equation}{0}
\def\theequation{A.\arabic{equation}}
\subsection*{Appendix A. Proof of steady state (\ref{P(C)}) to
  (\ref{f^a(x)})}

The steady state (\ref{P(C)}-\ref{f^a(x)}) can be demonstrated by
noting that, since the network model is defined in a fully connected
geometry (i.e.\ a link can be rewired from a node to any other node in
the network), the steady state probabilities must satisfy detailed
balance with respect to the rewiring dynamics. (In Section
\ref{generalisations} we comment on how one generalises to other
geometries.)  The detailed balance condition implies that
\begin{eqnarray} \label{balance}
u^s(n_{kl}{+}1)u^c(X_l{+}1) P(\{\underline{n}_l\};n_{kl}{+}1,n_{km}{-}1)  =
u^s(n_{km})u^c(X_m)  P(\{\underline{n}_l\}) \;,
\end{eqnarray}
where we have introduced the notation
$\{\underline{n}_l\};n_{kl}{+}1,n_{km}{-}1$ to represent a particle
configuration with $n_{kl}$ replaced by $n_{kl}{+}1$ and $n_{km}$
replaced by $n_{km}{-}1$. In this balance equation, the lhs represents
a particle hop from site $kl$ to site $km$ and the rhs represents the
reverse process. We now regard a solution of the form
(\ref{P(C)},\ref{f(n)}) as an ansatz, substitution of which into the
balance equation (\ref{balance})  yields
\begin{eqnarray}
u^s(n_{kl}{+}1) \frac{f^s(n_{kl}{+}1)}{f^s(n_{kl})} u^c(X_l{+}1)
  \frac{f^c(X_l{+}1)}{f^c(X_l)} = 
u^s(n_{km}) \frac{f^s(n_{km})}{f^s(n_{km}{-}1)}
  u^c(X_m) 
  \frac{f^c(X_m)}{f^c(X_m{-}1)}\;,
\end{eqnarray}
after cancelling common factors and rearranging slightly. This is
satisfied for 
\begin{equation}
u^a(x{+}1)f^a(x{+}1)=f^a(x)\;,
\end{equation} 
for $a=s$, $c$, which is iterated to give (\ref{f^a(x)}). This proves
the steady state (\ref{P(C)}) to (\ref{f^a(x)}).

\setcounter{equation}{0}
\def\theequation{B.\arabic{equation}}
\subsection*{Appendix B. Proof of steady state (\ref{P(C)gen}) to
  (\ref{f(n)gen})}   

The steady state (\ref{P(C)gen}) to
  (\ref{f(n)gen}) can be demonstrated by asking that the steady state
probabilities satisfy detailed balance with respect to the rewiring
dynamics, hence
\begin{eqnarray}
u_{kl}(\underline{n}_l; n_{kl}{+}1) \,
t_{kl'}(\underline{n}_{l'}; n_{kl'}{-}1)
P(\{\underline{n}_l\}; n_{kl}{+}1, n_{kl'}{-}1) =
u_{kl'}(\underline{n}_{l'})\,
t_{kl}(\underline{n}_l)  P(\{\underline{n}_l\}) \;, 
\end{eqnarray}
for each $k=1,\ldots,L$ and all pairs $l\neq l'$. Inserting the steady
state (\ref{P(C)gen}), cancelling common factors and rearranging
yields
\begin{eqnarray}
\frac{u_{kl}(\underline{n}_l; n_{kl}{+}1)}{t_{kl}(\underline{n}_l)}
\frac{f_l(\underline{n}_l; n_{kl}{+}1)}{f_l(\underline{n}_l)}  =
\frac{u_{kl'}(\underline{n}_{l'})}{t_{kl'}(\underline{n}_{l'}; n_{kl'}{-}1)}
\frac{f_{l'}(\underline{n}_{l'})}{f_{l'}(\underline{n}_{l'}; n_{kl'}{-}1)}\;,
\end{eqnarray}
again, for each $k=1,\ldots,L$. Both sides of this equation must be
equal to a constant which we choose, without loss of generality, to be
equal to one, whereby we obtain the recursion
\begin{equation} \label{recursion}
f_l(\underline{n}_l) =
\frac{t_{kl}(\underline{n}_l; n_{kl}{-}1)}
{u_{kl}(\underline{n}_l)}
f_l(\underline{n}_l; n_{kl}{-}1)\;, 
\end{equation}
which can be iterated to obtain (\ref{f(n)gen}). This recursion also
implies the constraint on the choice of the hop rates: one can obtain
two expressions for $f_l(\underline{n}_l)$ in terms of
$f_l(\underline{n}_l; n_{kl}{-}1, n_{k'l}{-}1)$; one comes by applying
the recursion (\ref{recursion}) first to $k$, then to $k'$; the other
comes by applying the recursion to $k'$ before $k$. Performing these
operations yields the constraint (\ref{constraint}).


\begin{thebibliography}{00} 

\bibitem{AB02}
R.~Albert and A.-L.~Barab\'asi, Rev.~Mod.~Phys.\ {\bf 47}, 74 (2002)

\bibitem{DM03}
S.~N.~Dorogovtsev and J.~F.~F.~Mendes 2003 
{\it Evolution of Networks} (OUP, Oxford) 

\bibitem{BA99}
A.-L.~Barab\'asi and R.~Albert,
Science {\bf 286}, 509 (1999) 

\bibitem{Newman01a}
M.~E.~J.~Newman,
    Proc.~Natl.~Acad.~Sci.~USA {\bf 98}, 404 (2001)

\bibitem{Newman04}
M.~E.~J.~Newman, Phys.~Rev.~E {\bf 70}, 056131 (2004)

\bibitem{GA04}
R.~Guimera and L.~A.~N.~Amaral, 
Eur.~Phys.~J. {\bf 38}, 381 (2004)

\bibitem{BBPV05}
M.~Barth\'elemy, A.~Barrat, R.~Pastor-Satorras and A.~Vespignani, 
    Physica~A {\bf 346}, 34 (2005) 

\bibitem{NR02}
J.~D.~Noh and H.~Rieger, 
Phys.~Rev.~E {\bf 66}, 066127 (2002)

\bibitem{MAB04}
P.~J.~Macdonald, E.~Almaas, A.-L.~Barab\'asi, 
cond-mat/0405688

\bibitem{YJBT01}
S.~H.~Yook, H.~Jeong, A.-L.~Barab\'asi and Y.~Tu,
Phys.~Rev.~Lett.\ {\bf 86}, 5835 (2001)

\bibitem{AK05}
T.~Antal and P.~L.~Krapivsky, 
    Phys.~Rev.~E {\bf 71},  026103 (2005) 

\bibitem{AKR05} E.~Almaas, P.~L.~Krapivsky and S.~Redner, Phys. Rev. E
{\bf 71}, 036124 (2005)

\bibitem{BBV04}
A.~Barrat, M.~Barth\'elemy and A.~Vespignani, 
    Phys.~Rev.~Lett.\ {\bf 92}, 228701 (2004)

\bibitem{BBV04a}
A.~Barrat, M.~Barth\'elemy and A.~Vespignani,
    Phys.~Rev.~E {\bf 70}, 066149 (2004)

\bibitem{BBV04b}
A.~Barrat, M.~Barth\'elemy and A.~Vespignani,
    Lect. Notes Comput. Sci. {\bf 56}, 3243 (2004)

\bibitem{DM04}
S.~N.~Dorogotsev and J.~F.~F.~Mendes,
cond-mat/0408343

\bibitem{BCK01}
Z.~Burda, J.~D.~Correia and A.~Krzywicki, 
Phys.~Rev.~E {\bf 64}, 046118 (2001) 

\bibitem{DMS03}
S.~N.~Dorogovtsev, J.~F.~F.~Mendes and A.~N.~Samukhin, 
Nucl.~Phys.~B {\bf 666}, 396 (2003)

\bibitem{AELM05}
A.~G.~Angel, M.~R.~Evans, E.~Levine and D.~Mukamel,
cond-mat/0503487

\bibitem{EH05}
M.~R.~Evans and T.~Hanney, J.~Phys.~A {\bf 38}, R195 (2005) 


\bibitem{KRL00}
P.~L.~Krapivsky, S.~Redner and F.~Leyvraz,
Phys.~Rev.~Lett.\ {\bf 85}, 4629 (2000)


\bibitem{KR01}
P.~L.~Krapivsky and S.~Redner, 
Phys.~Rev.~E {\bf 63}, 066123 (2001)  


\bibitem{BB01}
G. Bianconi and A.-L. Barab\'asi, 
Phys. Rev. Lett. {\bf 86}, 5632 (2001)


\bibitem{Evans00}
M.~R.~Evans, Braz.~J.~Phys.\ {\bf 30}, 42 (2000)                           

\bibitem{HJDXW05}
B.~Hu, X.-Y.~Jiang, J.-F.~Ding, Y.-B.~Xie and B.-H.~Wang,
cond-mat/0408125

\bibitem{EH03}
M.~R.~Evans and T.~Hanney, J.~Phys.~A {\bf 36}, L441 (2003)

\bibitem{HE04}
T.~Hanney and M.~R.~Evans, Phys.~Rev.~E {\bf 69}, 016107 (2004)

\bibitem{BGG03}
M.~Barth\'elemy, B.~Gondran and E.~Guichard, 
 Physica A {\bf 319}, 633 (2003)

\bibitem{B05}
G.~Bianconi, cond-mat/0412399

\bibitem{MBCV05}
A. De Montis, M. Barth\'elemy, A. Chessa and A. Vespignani,
cond-mat/0507106

\bibitem{GL02} 
C. Godr\`eche and J. M. Luck, J. Phys.:Condens. Matter {\bf 14}, 1601
(2002) 

\bibitem{EMZ}  
M. R. Evans, S. N. Majumdar and R. K. P. Zia 
J. Phys. A: Math. Gen. {\bf 37}, L275 (2004)

\bibitem{B03} 
M. Bal\'azs, Annales de l'Institut Henri Poincar\'e - PR {\bf 39}, 639
(2003)
 
\end{thebibliography}
\end{document}